\documentclass{article}

\usepackage{PRIMEarxiv}

\usepackage[utf8]{inputenc} 
\usepackage[T1]{fontenc}    
\usepackage{hyperref}       
\usepackage{url} 
\usepackage{ulem}

\newcommand*{\affaddr}[1]{#1} 
\newcommand*{\affmark}[1][*]{\textsuperscript{#1}}
\newcommand*{\email}[1]{\texttt{#1}}

\usepackage{indentfirst}
\usepackage{booktabs}       
\usepackage{amsfonts}       
\usepackage{nicefrac}       
\usepackage{microtype}      
\usepackage{lipsum}
\usepackage{fancyhdr}       
\usepackage{graphicx} 
\usepackage{multirow}
\usepackage{subfigure}
\usepackage{amsmath}
\usepackage{flushend}
\usepackage{color}

\graphicspath{{media/}}     
\begin{document}
\fancyhead[LO]{Q. Yang, Y. Zhao, H. Huang \textit{et al}}
\twocolumn[\begin{@twocolumnfalse}

\title{Fusing Blockchain and AI with Metaverse: A Survey
\thanks{ 
\textbf{Huawei Huang is the corresponding author.}} 
}
\author{%
Qinglin~Yang\affmark[1], Yetong~Zhao \affmark[1], Huawei~Huang\affmark[1*], Zehui~Xiong \affmark[2], Jiawen~Kang \affmark[3], and Zibin~Zheng\affmark[1]\\
\affaddr{\affmark[1]Sun Yat-sen University, China}\\
\affaddr{\affmark[2] Singapore University of Technology and Design, Singapore}\\
\affaddr{\affmark[3] Guangdong University of Technology, China}\\
\email{\{yangqlin6,huanghw28\}@mail.sysu.edu.cn}
}

\maketitle

\begin{abstract}
 \textit{Metaverse} as the latest buzzword has attracted great attention from both industry and academia. Metaverse seamlessly integrates the real world with the virtual world and allows avatars to carry out rich activities including creation, display, entertainment, social networking, and trading. Thus, it is promising to build an exciting digital world and to transform a better physical world through the exploration of the metaverse. In this survey, we dive into the metaverse by discussing how Blockchain and Artificial Intelligence (AI) fuse with it through investigating the state-of-the-art studies across the metaverse components, digital currencies, AI applications in the virtual world, and blockchain-empowered technologies. Further exploitation and interdisciplinary research on the fusion of AI and Blockchain towards metaverse will definitely require collaboration from both academia and industries. We wish that our survey can help researchers, engineers, and educators build an open, fair, and rational future metaverse.
 
\end{abstract}

\vspace{3mm}

\end{@twocolumnfalse}]
\footnotetext{Huawei Huang is the corresponding author.}

\section{Introduction}
\begin{figure*}[t]
    \centering
    \subfigure[Meeting collaboration.]{
        \centering
        \includegraphics[width=0.3\textwidth]{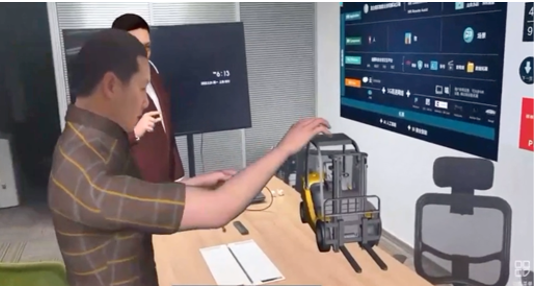}
         \label{fig:meeting}
    }
    \subfigure[Virtual exhibition.]{
        \centering
       \includegraphics[width=0.3\textwidth]{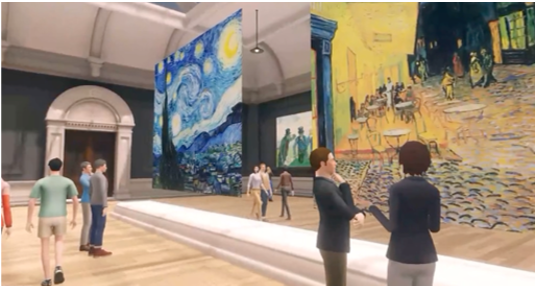}
         \label{fig:exhibition}

    }
        \subfigure[Metaverse concert \cite{virtualconcert}]{
        \centering
        \includegraphics[width=0.3\textwidth]{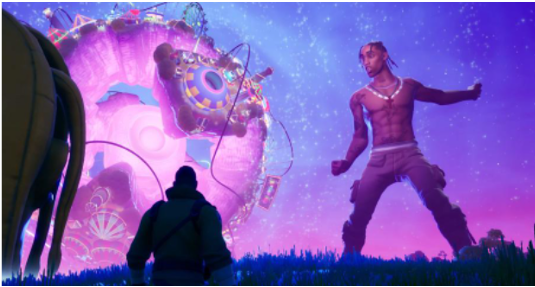}
         \label{fig:concert}

    }
    \quad
    
     \subfigure[Virtual and real symbiosis.]{
        \centering
        \includegraphics[width=0.3\textwidth]{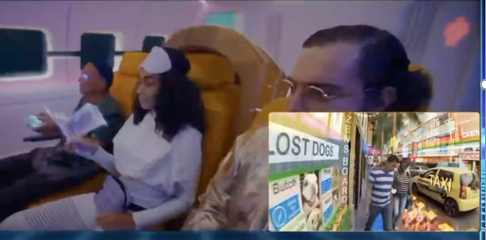}
         \label{fig:symbiosis}

    }
    \subfigure[$360^{\circ}$ Egocentric view of audience.]{
        \centering
     \includegraphics[width=0.3\textwidth]{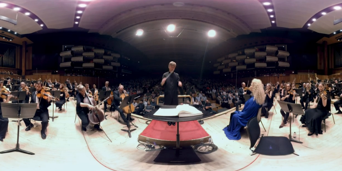}
         \label{fig:orchestra}
    }
        \subfigure[An animated virtual character's dance \cite{10.1145/3344383}.]{
        \centering
        \includegraphics[width=0.3\textwidth]{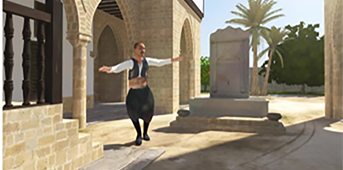}
         \label{fig:dance}
    }
    
    \caption{Effects that users can experience in metaverse. }
    \label{lbl:Technologies}
\end{figure*}

\begin{figure*}[t]
    \centering
    \includegraphics[width=0.8\textwidth]{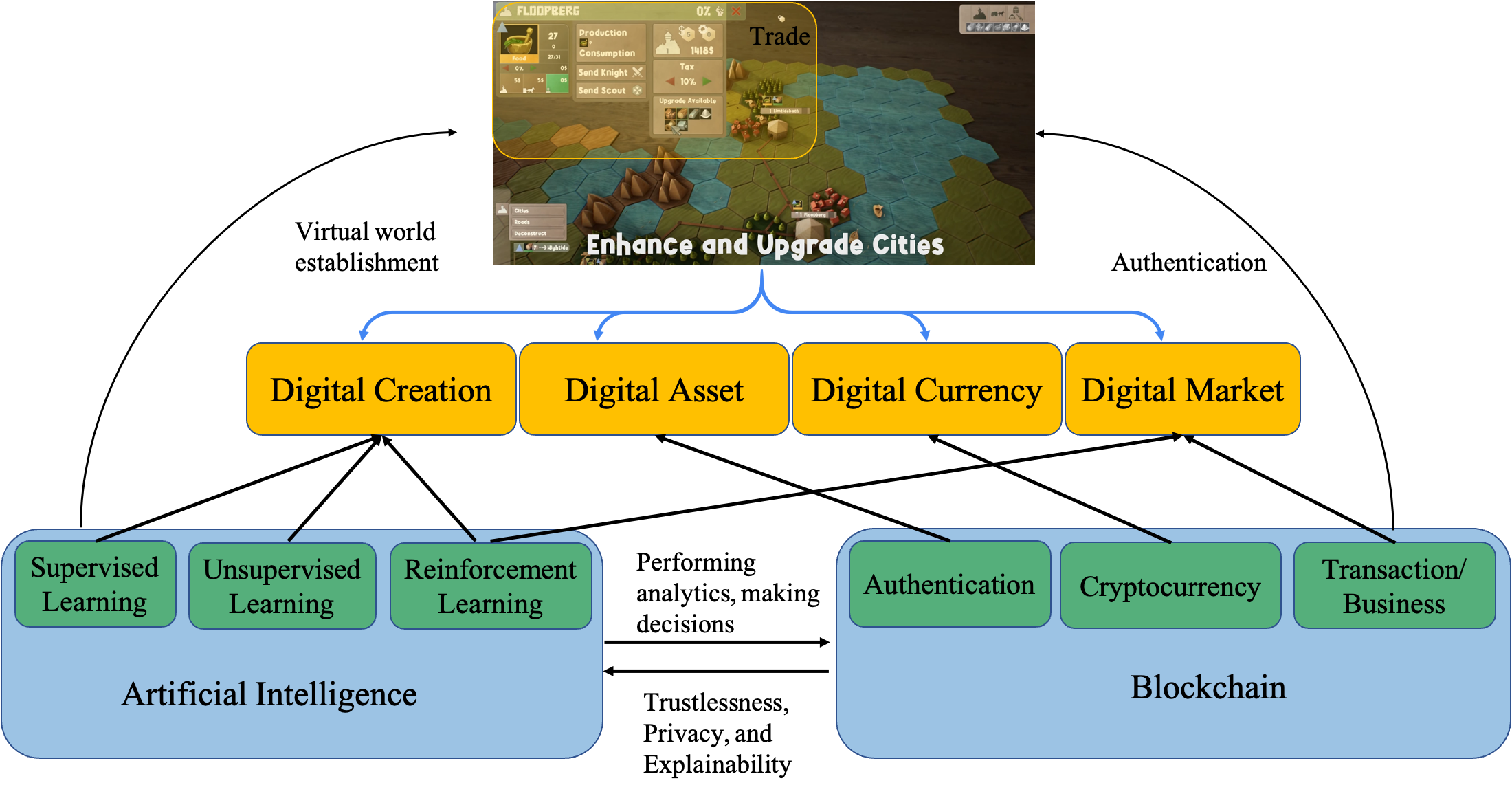}
    \caption{Fusion of AI and Blockchain in metaverse. Note that the picture at the top is from \url{https://store.steampowered.com/app/1313290/Let_Them_Trade/} }
    \label{fig:metaverse}
\end{figure*}

The concept of metaverse was proposed almost 30 years ago in the science fiction named \textit{Snow Crash}, written by \textit{Neal Stephenson}
\cite{joshua2017information}. Metaverse has been one of the hottest buzzwords to attract the tech industry’s attention due to the rapid advancements of Blockchain, Internet of Things (IoT), VR/AR, Artificial Intelligence (AI), Cloud/Edge Computing, {\color{black} etc}. The sandbox game platform Roblox \cite{roblox5} is the company that firstly incorporates the term `metaverse' into their prospectus and proposes the key characteristics (e.g., identity, friends, immersive experience, low friction, civility, economy, anywhere, variety) of metaverse. The social company, Facebook, is renamed as \textit{Meta} \cite{MetaLink} to help bring   metaverse to life and make people meet each other, learn, collaborate and play in ways that go beyond what they can imagine. The video game Fortnite \cite{games2017fortnite} that is released by Epic games puts the players into a virtual world, such as a post-apocalyptic, zombie-infested world, experiencing new levels of photorealistic interaction \cite{ambrasaite2021epic, Epic, seymour2017meet,games2017fortnite} and watching virtual concert \cite{virtualconcert}. Metaverse seamlessly integrates the physical world with the virtual world and allows avatars to carry out rich activities including creation, display, entertainment, social, and trading.  
{\color{black}
Nowadays, both academia and industries dive into the exploration of metaverse. 
For example, Jingteng Tech \cite{jingTech} has developed BeamLink, by which users can share photos, documents, and have an interactive meeting collaboration, as shown in Fig. \ref{fig:meeting}. Furthermore, the optical sensors (e.g., depth and cameras) deployed with CNNs are commonly applied to capture the gestures and movements of dancers. The poses and movements of dancers, as shown in Fig.\ref{fig:dance}, can be clustered as a high-dimensional feature space, which is further converted into dance
performance in virtual environments \cite{aristidou2019digital} through  techniques like \textit{dynamic time wrapping} \cite{ferguson2014dynamic}. 
Compared to the physical world, autonomous avatars can understand human dancer's voice-to-motion mapping. Powered by deep learning, avatars can also mimic dance moves with high dance resemblance and emotional expressions.
}
Given those potentials of metaverse, a problem is that researchers cannot accurately judge the shape and boundary of the future metaverse. They could only envision some of its possible characteristics, such as open space, decentralization, human-computer interaction experience, digital assets, and digital economy.

The avatars of human {\color{black}players}, their creations, and consumption in metaverse truly affect the physical world and even change the behaviors of people in the physical world, through the influence of people's thoughts (e.g., choosing the entertainment method). This change has a profound social significance \cite{likall,dionisio20133d}, and thus forms the {\color{black}lifestyle} of post-human society while reconstructing the digital economic system.
The metaverse can be viewed as a complete and self-consistent economic system, a complete chain of the production and consumption of digital items. 
The economy of metaverse refers to the digital production-based economic behaviors, e.g., creation, exchange, and consumption in the digital {\color{black}world are} the fundamental components of the digital economy.
The development of the economic system can be regarded as one of the most challenging tasks of   metaverse. This is because the production and consumption of digital assets that can be traded in the virtual world is a phenomenon that traditional economists have not encountered \cite{laduke1994traditional}.
Moreover, in a public, fair, and self-organized virtual world, the centralized economic system of the physical world cannot operate efficiently due to the high transaction volumes involved. Considering the interaction between the virtual world and the physical world,  metaverse should enable the currency circulation to break the barrier of life, products, learning, working, etc.
Therefore, the economic system of metaverse must be constructed in a decentralized manner such that the virtual assets of avatars could be traded efficiently in metaverse. 

Blockchain as a decentralized ledger without a centralized authority has drawn enormous attention in diverse application fields in recent years. Blockchain is highly expected to bring a variety of opportunities to metaverse, and trigger a new round of technological innovation and industrial transformation. 
On the other hand, recent advances in AI have brought promising solutions to overcoming the challenges of metaverse development, such as big data analytics, AI-empowered content generation, and intelligence deployment. Consequently, the integration of AI and blockchain becomes a promising trend to promote the benign evolution of the blockchain/AI-empowered metaverse ecosystem.
Although the advent of blockchain and AI has spawned a large number of new technologies and applications, the fusion of blockchain and AI with metaverse also poses several emerging research challenges. For example, transaction volumes in metaverse {\color{black}systems} are much higher than those in the physical world due to the features of digital products and markets. Blockchain-based Non-Fungible Tokens (NFT) enable avatars to generate the content that can be traded with their digital certificates \cite{nadini2021mapping,lambert2021beyond}. 

We then review several representative survey articles here, to highlight the difference between our survey.
Lim \textit{et al.} \cite{lim2022realizing} focus on the network demands \cite{xu2021wireless} of metaverse from the perspective of edge intelligence \cite{ng2021unified} since metaverse is viewed as `the successor to the mobile Internet'.
Du \textit{et al.} \cite{DBLP:journals/corr/abs-2111-00511} propose a privacy-preserving targeted advertising strategy for the wireless edge metaverse to enable metaverse service providers to allocate network bandwidth to users so that the users can access metaverse from edge sites.
Jiang \textit{et al.} \cite{DBLP:journals/corr/abs-2111-10548} introduce a kind of collaborative computing paradigm based on Coded Distributed Computing to support the computation requirement of metaverse services \cite{han2021dynamic}.
The up-to-date survey \cite{likall} mainly reviews the state-of-the-art technologies as enablers to implement metaverse, such as high-speed networks (e.g., 5G) and edge computing, Blockchain, and AI. 
The authors' findings demonstrate the gap between the up-to-date technologies and the demands of implementing metaverse. 
The other survey \cite{lee2021study} focuses on metaverse analytics, the search traffic, news frequency, and the topic concerning sustainable growth. 
Duan \textit{et al.} \cite{duan2021metaverse} highlight the representative applications for social goods and propose a three-layer metaverse architecture from a macro perspective, which contains infrastructure, interaction, and ecosystem.
In contrast, our survey discusses how to fuse blockchain and AI technologies with metaverse.
As depicted in Fig. \ref{fig:metaverse}, AI technologies are applied to the digital creation and digital market. Meanwhile, blockchain can guarantee digital assets, digital currencies, and the digital market.
On the other hand, in this survey, we emphasize the fusion of AI and blockchain technologies to establish an intelligent, open, fair, and promising future metaverse.

The contributions of the paper include the following three points.
\begin{itemize}

    \item We first present the preliminaries of the economic system in the metaverse.
    
    \item We then discuss how blockchain and AI technologies fuse with the metaverse, and review the state-of-the-art studies.

    \item Finally, we envision typical challenges and open issues to shape the future metaverse in the next decades.
    
\end{itemize}

The rest of this paper is organized as follows. Section \ref{sec:Economy} mainly describes the characteristic of the ecosystem of metaverse by comparing it with the conventional economy. Section \ref{sec:AI} discusses the artificial intelligence technologies and the challenges of applications in the metaverse. In Section \ref{sec:blockchain}, the fundamental Blockchain technologies and applications are introduced in metaverse by comparing with the current breakthroughs. Section \ref{sec:challenges} mainly discusses the challenges and open issues of shaping the future metaverse. Finally, Section \ref{sec:conlusion} concludes this paper.
\section{Preliminaries of Economic System in Metaverse}\label{sec:Economy}
\begin{figure}
    \centering
    \includegraphics[width=0.5\textwidth]{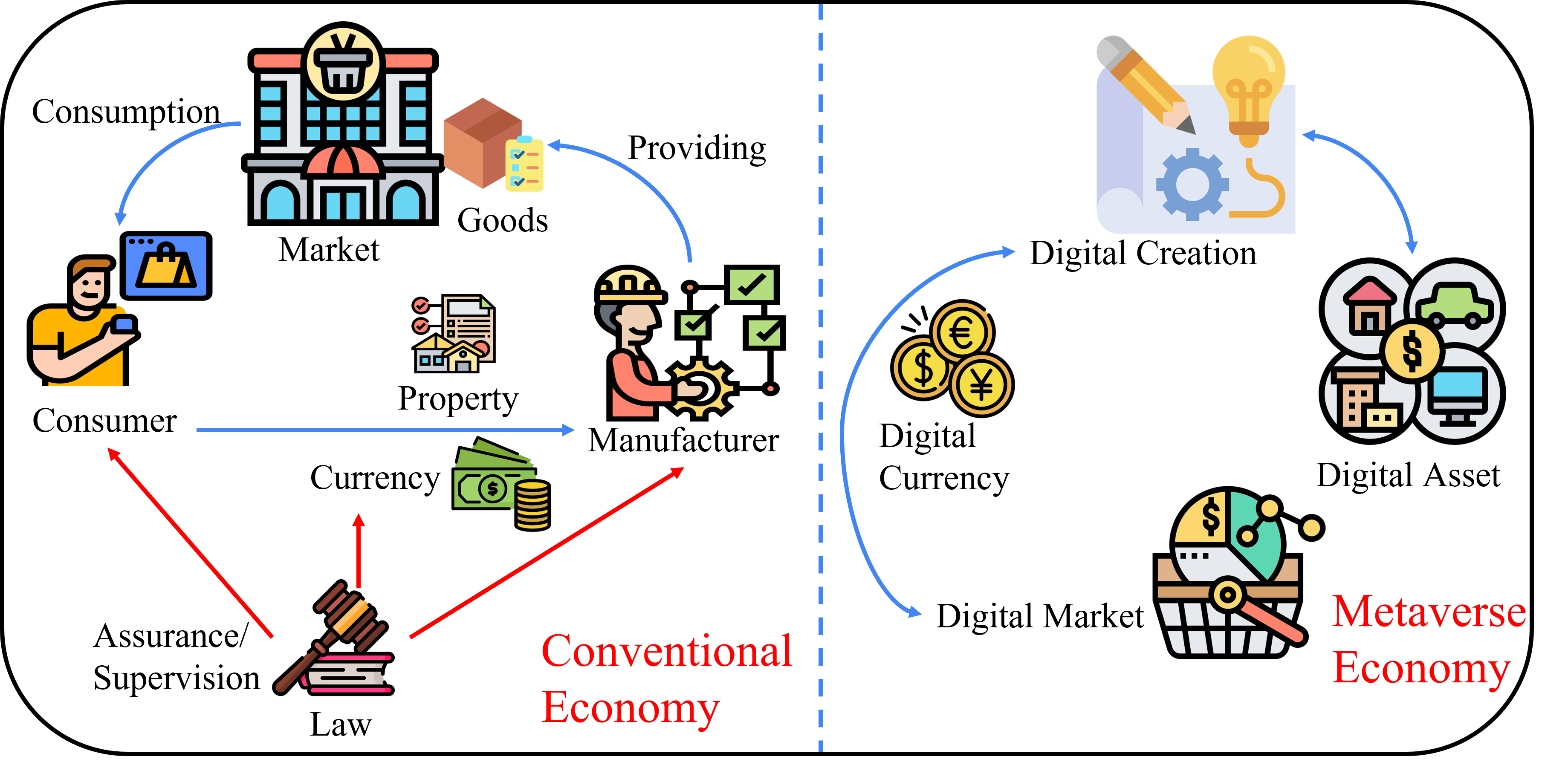}
    \caption{Conventional economy and metaverse economy. icons are from \url{https://www.flaticon.com/}}
    \label{fig:economy}
\end{figure}

The economy is the fundamental component of the metaverse. From the more idealistic perspective, metaverse should be interoperable such that users can trade virtual items like clothes or cars from one platform to another. 
Firstly, as depicted in Fig. \ref{fig:economy}, we shall describe the metaverse economic system according to the mainstream games and existing research works. The metaverse economic system is composed of four parts: digital creation, digital asset, digital market, and digital currency, whose exploitation will lead to the transformation of the conventional economy. 
\begin{itemize}
   \item \textbf{Digital Creation} is the foundation of the metaverse. The creation progress is similar to the material products in the physical world.
   The development of the economy of the metaverse is decided by the number of creators. Thus, the digital creation activities urgently need a basic authoring tool that can make the creation easily and personalized \cite{rymaszewski2007second, arnaud2006collada}.
  The Decentraland \cite{decentraladCreation} is a game platform that provides two kinds of authoring tools for creating interactive Decentraland applications, i.e., the \textit{Builder} and the \textit{Decentraland SDK}.
  Players can execute a simple drag and drop editor by the \textit{Builder} without coding required, while the \textit{Decentraland SDK} enables players to have sufficient freedom to create their applications. 
  Lee \textit{et al.} \cite{lee2021creators} summarize the research works on computational arts that are relevant to the metaverse, describing novel art work (e.g., immersive arts, robotic arts.) in blended virtual-physical realities. 
   \item \textbf{Digital Asset} has the hidden property, which is the precondition of trade. 
   For instance, in the FPS game \textit{Counter-Strike: Global Offensive}, players can equip their weapons with all kinds of `skins' that shows the asset attributes the `skins' since the `skins' can be exchanged, traded, or bought at the platforms (e.g., BUFF, Steam). 
  {\color{black}However, }these trade platforms need to make the users' accounts known to the public, which might incur privacy issues since there is no reliable mechanism to guarantee the digital asset and platform.
 Hence, people's confirmation of digital assets is inseparable from the value provided by the blockchain and the encryption system of the blockchain. 
 Because encryption algorithms can capitalize on data, and consensus mechanisms help people verify and confirm transactions.
   \item \textbf{Digital Market} is the fundamental place in which avatars can trade to have income like in the physical world. 
   The mature market of metaverse that should ensure the creation of products and real trade accomplished in metaverse must be different from the existing digital market. 
   Bourlakis \textit{et al.} \cite{bourlakis2009retail} examine the evolution of retailing, i.e., from traditional to electronic to metaverse retailing, and sheds light on the ways metaverse influence that evolution.
 {\color{black}Cliff \textit{et al.} \cite{9308172} mainly discuss the research that applies computational intelligence, i.e., the methods from AI and machine learning to automatically discover, implement, and fine-tune strategies for adaptive automated trading in financial markets.}
 Decentraland marketplace \cite{decentralandmarket} allows users to trade all their Decentraland on-chain assets like what they could behave in the physical world.
   \item \textbf{Digital Currency} is the media in metaverse with which the avatars can finish the trade and exchange. 
   While in the metaverse, fiat currency cannot satisfy the demands of the development of the metaverse due to the high cost of the legal currency system. 
   In addition, fiat currency is converted from physical currency (e.g., gold, silver) that is different from the digital-based currency of metaverse. 
   Roblox enables players to buy various items by purchase with real currency, from a recurring stipend given to members with the premium membership, or from other players by producing and selling virtual content in Roblox \cite{roblox5}.
   The Diem payment system \cite{diem} supports single-currency stablecoins (e.g., USD, EUR, and GBP) and a multi-currency coin (XDX). We shall talk about the fusion of blockchain in metaverse in section \ref{sec:blockchain}.
\end{itemize}

As demonstrated in \textit{The Wealth of Nations},  the traditional economics in the real world are based on some basic premises such as the scarcity of resources, and selfish individuals \cite{smith1937wealth}. 
In contrast, the individuals in the metaverse are selfless and illogical. These individuals prefer to emphasize their personal feeling such as happiness and a sense of accomplishment. This is because there is no farming society and industrial society experienced by human beings, and no traditional industrial structure in the virtual world.
{\color{black}In the metaverse, the conceptual economy will be the basic form of economic activities. The natural form of financial currency can no longer be precious metals, but the virtual social currency. }
Hence, the difference between the economy of the metaverse and the conventional economy can be summarised as follows:
\begin{itemize}
\item In metaverse, identity determines value instead of the undifferentiated labor in the conventional economy.
\item The marginal benefits will increase in metaverse instead of diminishing marginal benefits of production in the physical world.
\item The marginal costs of products will decrease, compared with the physical world. 
\item Transaction costs in metaverse tend to zero, which will incur frequent transactions.
\end{itemize}

 Now, the metaverse economic system is still in its early stage, because it can barely transplant and test the innovations of the digital economy, including various digital currencies, test cooperative economy, sharing economy, and inclusive finance. 
 {\color{black}
 To enable an intelligent and secure metaverse, we think that key technologies such as AI and blockchain should be fused with the metaverse. Therefore, we review the related studies that can offer us inspiration in the following sections.
 }
 
\section{Artificial Intelligence in Metaverse} \label{sec:AI}
\begin{figure*}[htbp]
    \centering
    \subfigure[Support-Vector Machine]{
        \centering
        \includegraphics[width=0.28\textwidth]{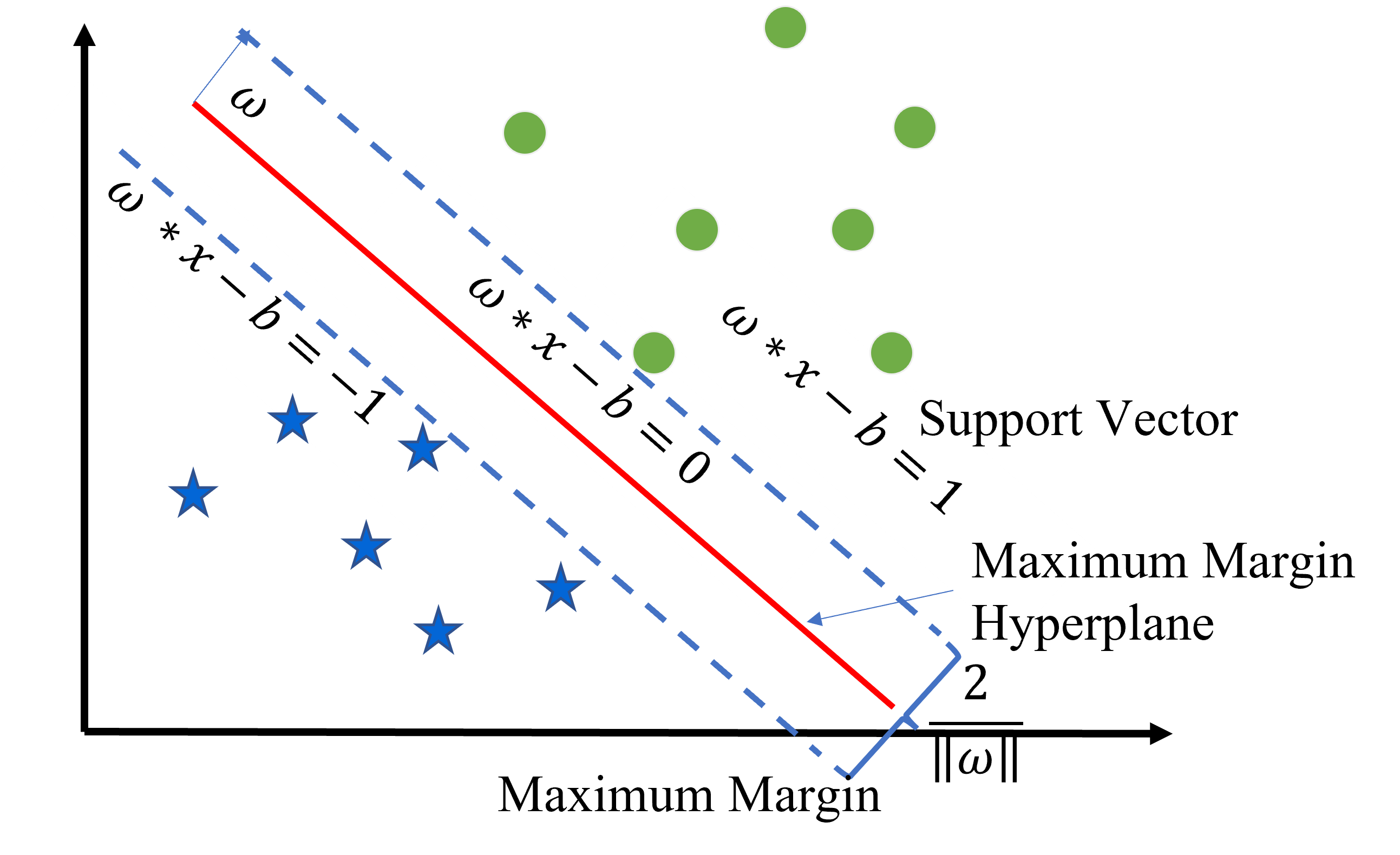}
        \label{lbl:svm}
    }
    \subfigure[Convolutional Neural Network]{
        \centering
        \includegraphics[width=0.38\textwidth]{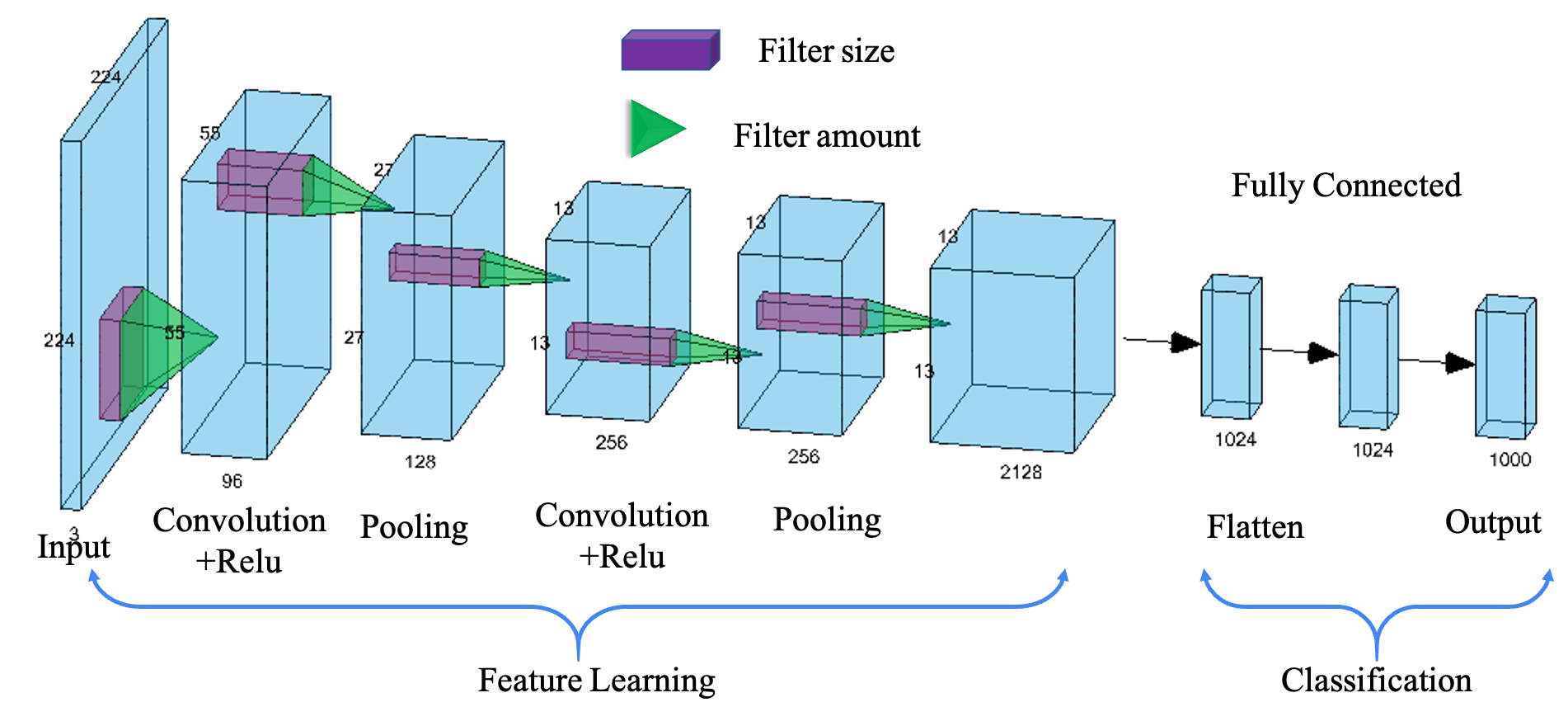}
        \label{lbl:cnn}
    }
        \subfigure[Reinforcement Learning]{
        \centering
        \includegraphics[width=0.28\textwidth]{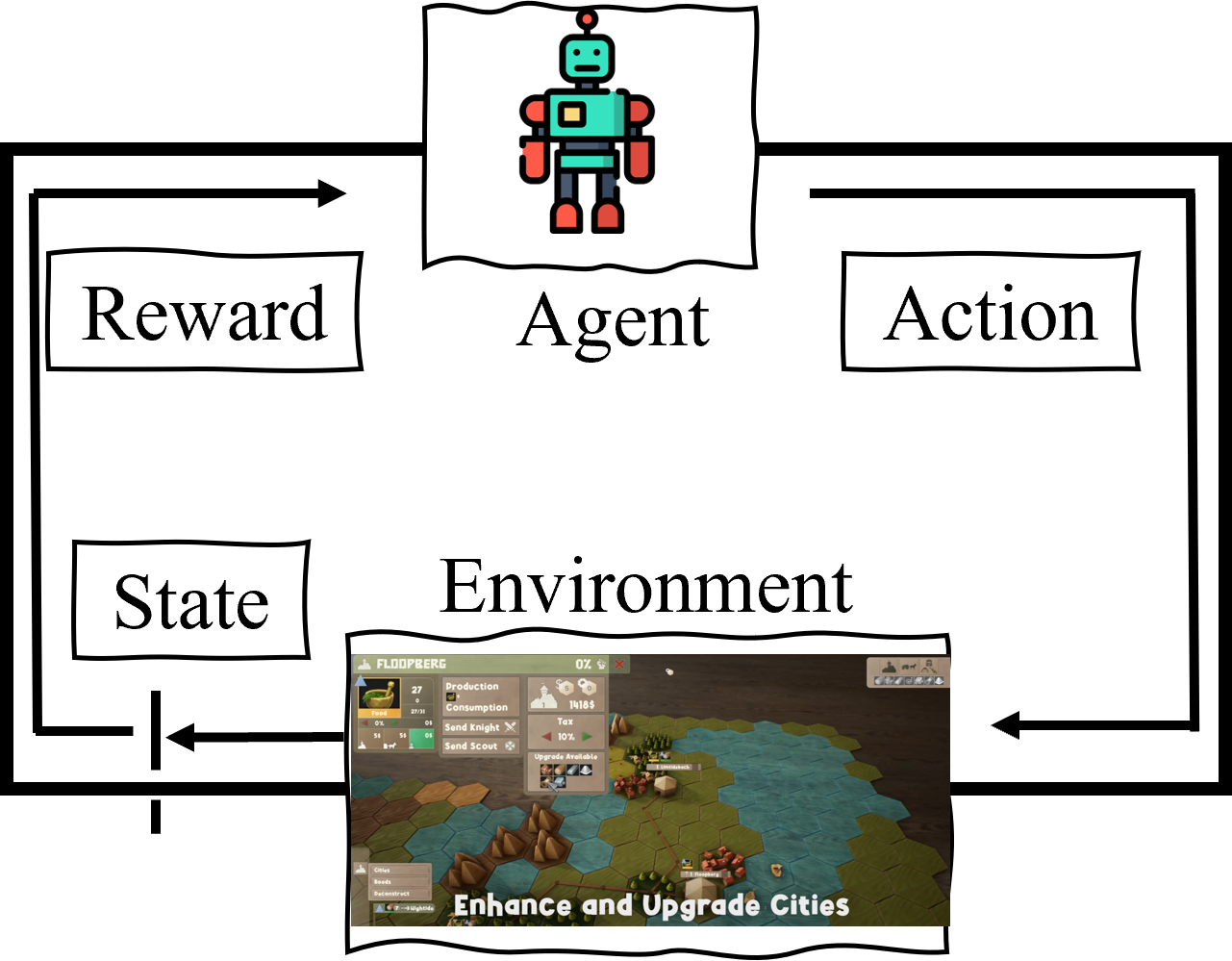}
        \label{lbl:RL}
    }
    \caption{ Illustration of artificial intelligence technologies }
    \label{lbl:Technologies}
\end{figure*}
{\color{black} Artificial Intelligence is a research discipline proceeded based on the hypothesis that every aspect of learning can in principle be so precisely described} \cite{dick2019artificial}. 
The state-of-the-art AI studies focus on machine learning, deep learning, and reinforcement learning in the fields including computer vision, decision-making, {\color{black}and} natural language process (NLP).
Intuitively, the breakthroughs of artificial intelligence in the real world motivate people to realize metaverse. 
{\color{black}
 For example, machine learning provides technical support for all systems in metaverse to reach or exceed the level of human's learning. 
 It shall significantly affect the operational efficiency and the intelligence of metaverse.
 Intelligent voice services provide technical support, such as voice recognition and communication, for metaverse users.
 }
\subsection{Representative AI Algorithms}
Machine learning algorithms (e.g., linear regression \cite{jammalamadaka2003introduction}, random forest \cite{oshiro2012many}, singular value decomposition \cite{paige1981towards}) enable machine to have the human ability by learning from experience and data. 
For example, support vector machine (Fig. \ref{lbl:svm}) \cite{vapnik1999nature} is a kind of representative machine learning algorithm and is used for the problem of pattern classification, regression, and learning a ranking function.
The support vector classification aims to find an optimization hyperplane to separate the dataset by minimizing the following object function:
\begin{equation}
L(\omega)=\sum_{i=1} \underbrace{\max \left(0,1-y_{i}\left[\omega^{T} x_{i}+b\right.\right.}_{\text {Loss function }}] \underbrace{+\lambda\|\omega\|_{2}^{2}}_{\text {regularization }},
\end{equation}
where $\omega$ denotes a weight vector, $b$ denotes the threshold, and $\lambda$ denotes a Lagrangian factor that determines the trade-off between margin maximization and regularization in the loss function. 
However, machine learning algorithms usually require to select features manually, which limits its wide applications since a large amount of labeled data is needed.

The convolutional neural networks (CNNs, or ConvNets (Fig. \ref{lbl:cnn})) are a kind of representative deep neural network inspired by biological neural network. 
The normal CNNs are based on the shared-weight architecture of the convolution kernels or filters that slide along input features and provide translation equivariant responses known as feature maps. 
CNNs are always composed of convolution layers, pooling layers and fully connected layers \cite{ketkar2017convolutional}. 
The great amount of reduction in CNN is achieved by a technique called weight sharing between neurons. Given an image $X \in \mathbb{R}^{M \times N}$ and a filter $W \in \mathbb{R}^{m \times n}$, where $m << M$, $n << N$, the convolution can be written follows:
\begin{equation}
    y_{ij}=\sum_{u=1}^{m}\sum_{v=1}^{n}\omega_{uv}\cdot x_{i-u+1,y-v+1},
\end{equation}
where $\omega$ is the shared training parameters.
Thus, CNNs are regarded as a kind of prevalent supervised learning that could perform well in many computer vision applications such as facial recognition, image search, augmented reality, and more.

While reinforcement learning describes the sequential decision-making problem faced by an agent that must learn experience through trial-and-error by interacting with a dynamic environment \cite{kaelbling1996reinforcement}. The schematic of RL is demonstrated in Fig. \ref{lbl:RL}.
{\color{black}Considering} the Markov decision processes (MDPs) \cite{van1980stochastic} and deep neural network, deep reinforcement learning is promising to revolutionize the field of AI and represent a step towards establishing an autonomous system with a higher level of understanding of the visual world \cite{arulkumaran2017deep}. 
 And there are two main approaches to solving RL problems: value functions and policy search. 
Value function methods are based on estimating the value (expected return) of being in a given state. The optimal policy $\pi^{*}$, has a corresponding state-value function $V^{*}(s)$, and vice versa; the optimal state-value function can be defined as:
\begin{equation*} V^{\ast}(\mathbf{s})=\max_{\pi}\mathbb{E}[R\vert \mathbf{s},\ \pi]. \ \forall \mathbf{s}\in \mathcal{S}, \tag{3} 
\end{equation*}
where, $s$ denotes the state, and $\pi$ denotes the policy that agent follows.
By contrast, policy search methods do not need to maintain a value function model but directly search for an optimal policy $\pi^{*}$.
Until now, there have been developed many deep learning-based RL (DRL) algorithms, including deep Q-network (DQN), trust region policy optimization (TRPO), and asynchronous advantage actor-critic. These DRL algorithms can be applied to achieve over-human performance in some fields \cite{8103164}.
In the next section, we review the works on AI technologies that are related to the metaverse when it comes to the establishment of a metaverse environment and object creation in the metaverse.

\subsection{Establishment of Metaverse Environment}
Not only metaverse users, but the objects or things in the physical world also interact with the metaverse, evolving to persistently represent the structure, behaviors, and context of a unique physical asset (such as a component, a human, or a process) \cite{kapteyn2021probabilistic} in the virtual world. 
With the breakthroughs of digital transformation, the latest trend in every industry is to build digital twins with the ultimate goal of using them throughout the whole asset life-cycle with real-time data \cite{san2021digital}.
Digital twins are instrumental not only during the conceptualization, prototyping, testing, and design optimization phase but also during the operational phase.
The virtual world of metaverse generates a huge amount, variety, and velocity of data, such as structured data, and unstructured data, which makes deep learning-based digital twin (DT) essential \cite{qi2018digital}. 
It can provide a better understanding of the underlying mechanics to all the stakeholders by the fusion of the virtual world with data sciences \cite{malik2017merging}.
Ham \textit{et al.} \cite{ham2020participatory} propose a new participatory-sensing-to-digital twin city framework for community functioning in cities. The work fuses crowdsourced and unstructured visual data-based reality information with a three-dimensional (3D) virtual city to update the 3D city model that is fed into a computer-aided virtual environment (CAVE) for interacting and immersive visualization. 
To address the challenge of processing unstructured point clouds, epitomized by high cost, movable objects, limited object classes, and high information inadequacy/redundancy.
Xue \textit{et al.} \cite{xue2020lidar} present a novel unsupervised method, called Clustering Of Symmetric Cross-sections of Objects (COSCO), to process urban LiDAR point clouds to a hierarchy of objects based on their characteristic cross-sections. COSCO follows the Gestalt design principles, including proximity, connectivity, symmetry, and similarity. 
 
\begin{table*}
\caption{A Summary of the Application of AI in Metaverse}
\center
\footnotesize
\setlength\tabcolsep{1.5pt} 

 \begin{tabular}{|p{2.5cm}|p{5.0cm}|p{5.0cm}|p{5cm}|}
\hline
\bf{ Types} & \bf{Description} & \bf{Machine Leaning Models}  & \bf{Use cases}  \\ \hline \hline
\multirow{7}{*}{\shortstack[l]{Virtual \\ Environments }}
&  3D computer vision \cite{lai2018vivid} & DRL & Learning indoor navigation, Action recognition, Event detection, etc.  \\ \cline{2-4}
& Federated learning \cite{9049708}& Parameter server-based  & Augmented reality applications \\ \cline{2-4}
& To reduce the executing latency and the drawbacks of AR \cite{9049708}  & Centralized FL in mobile edge computing  & Collaborative learning \\ \cline{2-4}

& Enabling Cognitive Smart Cities Using Big Data and Machine Learning  \cite{8291121}  & Semi-supervised deep reinforcement learning  & Smart city services \\ 

\hline

\multirow{4}{*}{\shortstack[l]{}} &  Recognizing Avatar Faces \cite{6406586}&Markov random field  &  Face Recognition  \\ \cline{2-4}
&Detection and track \cite{fabbri2018learning} & Three-branch multi-stage CNN (Fig. \ref{lbl:cnn}) &Multi-people tracking \\ \cline{2-4}
&NPC training \cite{wang2009rl}  & RL (Fig. \ref{lbl:RL}) &RL-DOT \\ \cline{2-4}
& OpenAI Five \cite{berner2019dota}
& Distributed learning framework and LSTM  & Dota2 \\ 
\cline{2-4}

\multirow{1}{*}{\shortstack[l]{AI-based Object}} 
&Intelligent behavior avatar \cite{1506510}  & RL-based bayesian networks graph &Play game  tracking \\ \cline{2-4}
&Learning-based interactive avatar control \cite{lee2006precomputing}& State-action & Animate and control avatars \\ \cline{2-4}
&Human-computer interaction \cite{kastanis2012reinforcement}& RL (Fig. \ref{lbl:RL}) &Avatar moving  \\ 
\hline 

\multirow{2}{*}{\shortstack[l]{Virtuality-Reality \\ Interaction}} 
& The trained controller in virtual environment can be transferred to the physical world \cite{rahmatizadeh2018virtual,rahmatizadeh2016learning,wang2021digital} & LSTM and Mixture density network & Robots training, Digital twin for human-machine interaction \\ \cline{2-4}
\hline 
\end{tabular}
\label{lbl:AIApplication}
\end{table*}

Lai \textit{et al.} \cite{lai2018vivid} propose a novel virtual environment for visual deep learning since the existing works have the drawbacks, such as small scenes or limited interactions with objects, which provides large-scale diversified indoor and outdoor scenes.
Augmented reality (AR) devices could provide people with immersive and interactive experiences while their applications are latency-sensitive. Hence, in the work \cite{9049708}, Chen \textit{et al.} exploit to address the computation efficiency, low-latency objects recognition, and classification issues of AR applications by integrating mobile edge computing paradigm with federated learning. 
A city DT system depends on long-term and high-quality data to bring to perfection. Pang \textit{et al.} \cite{pang2021collaborative} propose a federated learning-based DT framework to enable multiple city DTs to share the local strategy and status quickly while accumulating the insights from multiple data sources efficiently, thereby enhancing privacy protection settings. 

Lee \textit{et al.} \cite{lee2021collaborative} present a serious game for self-training fire evacuation drills, in which the avatar is synchronized with multiple trainees and can be placed in different remote physical locations with the option of real-time supervision.
The proposed system architecture includes a wearable motion sensor and a head-mounted display to synchronize each user's expected motion with her/his avatar activity in the cyberspace of the metaverse environment. This system architecture provides an immersive and inexpensive environment for the easy-to-use user interface of a fire evacuation training system based on network experience. 
The model \cite{fabbri2018learning} explicitly deals with occluded body parts by hallucinating plausible solutions of not visible joint. Fabbri \textit{et al.} propose a new end-to-end architecture that is a three-branch multi-stage CNN with four branches (visible heat-maps, occluded heat-maps, part affinity fields, and temporal affinity fields) fed by a time linker feature extractor. 

To overcome the lack of surveillance data with tracking, body part, and occlusion annotations they created the vastest computer graphics dataset for people tracking in urban scenarios by exploiting a photorealistic video game.
During the initial stage of the metaverse, it still requires technological companies or governments to collaboratively establish the AI-based infrastructure when it comes to the computational capacities, data, technologies, etc. 
While scattered ownership of data is another barrier since companies often don’t want to share commercially sensitive information, nor do governments \cite{tao2019make}. 
To address this issue, federated learning (FL) has emerged as a kind of collaborative learning paradigm, allowing participants to train the shared model locally by transferring the training parameters instead of raw data. 
FL paradigm can protect the data privacy and reduce the communication overhead \cite{li2020federated}, especially for the large-scale scenarios with large {\color{black}models} and massive data. 
With respect to the data privacy, there have been many research on applying FL in medical institutions \cite{rieke2020future}, industries \cite{tao2019make}, banks \cite{yang2019federated}, etc.

Mohammadi \textit{et al.} \cite{8291121} propose a semi-supervised deep reinforcement learning-based framework that utilizes a fusion of labeled (users' feedback) and unlabeled (without such users' feedback) data to converge toward better control policies instead of wasting the unlabeled data. The proposed framework is scalable to satisfy the demands of smart city services. Intuitively, this research on semi-supervised deep reinforcement learning-based can be mapped into the service of the metaverse. While the development of intelligent metaverse remains the challenges, such as integrating big and fast/streaming data analytics, big dataset shortage, {\color{black}and} on-device intelligence.

\subsection{Object Creation in Metaverse}
After the descriptions of AI-based establishment of virtual world, we shall argue the authoring tools in metaverse since AI-based authoring tools provide technical support for all systems and roles to reach or exceed the level of human learning. Authoring tools will greatly affect the operational efficiency and intelligence of metaverse.

\subsubsection{Avatar and Non-player Characters}

The notion \textit{avatar} can be derived from the Sanskrit word. It identifies the god Vishnu’s manifestations on earth. However, it was the first to be used for player representations in virtual worlds \cite{castronova2004price}. 
Avatars are not only used in games, but also as users’ representations in e-commerce applications, virtual social environments, and in geographically separated workplace meetings \cite{schroeder2012social}. 

Chen \textit{et al.} \cite{1506510} propose a novel method for personal intelligent behavior avatar to make the optimal strategic decision for the user through the interactions between the user and agents by integrating Bayesian Networks and reinforcement learning in the virtual environment. 
To create a controllable and responsive avatar with large motion sets in computer games and virtual environments, Lee \textit{et al.} \cite{lee2006precomputing} presents a novel method of precomputing avatar behavior from unlabeled motion data in order to animate and control avatars at minimal runtime cost. 
Meanwhile, a reinforcement learning method \cite{kastanis2012reinforcement} is applied to train a virtual character to move participants to a specified location.
The virtual environment demonstrates an alleyway displayed through a wide field-of-view head-tracked stereo head-mounted display.
This method opens up the door for many such applications where the virtual environment adapts to the responses of the human participants with the aim of achieving particular goals.

Apart from the mentioned above, AI-driven non-player characters (NPCs) are computer-operated characters who act as enemies, partners, and support characters to provide challenges, offer assistance, and support the storyline.
While from the game perspective, most of the human-looking NPCs are not intelligent enough to interact with players in specific game genres (e.g., real-time strategy, some modes of first-person shooting), which requires strong tactical decision-making abilities.
Wang \textit{et al.} \cite{wang2009rl} propose a reinforcement learning-based domination team for playing unreal tournament (UT) domination games that consists of a commander NPC and several solider NPCs. 
During each decision cycle in the running process, the commander NPC {\color{black}decides} troop distribution and, according to that decision, sends action orders to other soldier NPCs. 
Each soldier NPC tries to accomplish its task in a goal-directed way, i.e., decomposing the final ultimate task (attacking or defending a domination point) into basic actions (such as running and shooting) that are directly supported by UT application programming interfaces (APIs).
The RL agents as means of creating NPCs that could both progressively evolve behavioral patterns and adapt to the dynamic world by exploring their environment and learning optimal behaviors from interesting experiences \cite{merrick2006motivated,razzaq2018zombies}. 

Berner \textit{et al.} \cite{berner2019dota} develop a distributed training system and tools for continual training which allows researchers to train OpenAI Five for 10 months. By defeating the Dota 2 \cite{dotaGuide} world champion (Team OG), OpenAI Five demonstrates that self-play reinforcement learning can achieve superhuman performance on a difficult task.
 Rahmatizadeh \textit{et al.} \cite{rahmatizadeh2018virtual} attempt to address the challenging problem of behavior transfer from virtual demonstration to a physical robot through training a Long Short Term Memory (LSTM) recurrent neural network to generate trajectories.
 During the training process, a Mixture Density Network (MDN) is applied to calculate an error signal suitable for the multimodal nature of demonstrations. 
 The learned controller in the virtual environment can be transferred to a physical robot (a Rethink Robotics Baxter) and successfully perform the manipulation tasks on a physical robot, which motivates the avatar to create AI objects that can impact the physical world.
 Similarly, the works \cite{rahmatizadeh2016learning,wang2021digital} exploit the interaction between the virtual environment and physical world by CNNs.

\subsubsection{AI-based Activities in Metaverse}

In games (e.g., \cite{Epic}, \cite{roblox5}, \cite{decentralandIntro}), the basic characteristics of metaverse can be perfectly explained and displayed. 
{\color{black}
However, no game has fully achieved} an ideal metaverse.
 {\color{black} Games conventionally have rules, objectives, and boundaries that enable them to shape a specific gameplay. In contrast, metaverse does not require any specific gameplay.
 Some online social games are very similar to metaverse, such as \textit{sims} \cite{decentralandIntro}.
 Metaverse, however, differs from video games, because it involves many activities that are not necessarily for fun. Examples are reviewed as follows.
\begin{itemize}
    \item Metaverse users can attend events (concert, virtual exhibition, remote education, meeting collaboration, etc.) without having to travel.
    
    \item Metaverse is virtual and real symbiosis, which means it can evolve in parallel even if people leave the virtual world anytime.
\end{itemize}}

Being aware of the difference between video games and metaverse, we can exploit metaverse from the perspective of video games, and extend it to the fields of manufacturing, education, creation, entertainment, social, and so on.
Ando \textit{et al.} \cite{ando2012level} present a way how to infer the observed exhibits in a metaverse museum from a movement log based on Second Life. To use recommendation systems in metaverse museums, they need some pieces of information to infer which exhibits the user is visiting via performing this task efficiently and precisely by focusing on the avatar’s states in the museum. 
%
%

Yampolskiy \textit{et al.} \cite{6406586} propose a set of algorithms that are capable of verification and recognition of avatar faces with a high degree of accuracy.
 Lugrin \textit{et al.} \cite{lugrin2006ai} propose a method for the AI-based simulation of object behavior so that interactive narrative can feature the physical environment inhabited by the player character as an `actor'. 
The prototype based on the top of the Unreal Tournament game engine relies on a `causal engine', which essentially bypasses the native Physics engine to generate alternative consequences to player interventions. 
The evaluation method \cite{kreminski2019evaluating} can be applied to the human-centered evaluation of AI-based games, grounded in the analysis of player retellings of their play experiences in Civilization VI, Stellaris, and two distinct versions of the research game Prom Week.
The reason is that it is difficult to understand through existing evaluation methods, such as the typical narrative structure that players tend to have in their minds when playing a specific game. The diversity of subjective experience narratives that might occur in a specific game.

Puder \textit{et al.} \cite{puder1995ai} demonstrate that an open distributed environment can be viewed as a service market where services are freely offered and requested. 
Any infrastructure which pursues appropriate mechanisms for such an environment should contain mediator functionality (i.e., a trader) that matches service demands and service offers. 
%
%
%
%

In the open and decentralized metaverse, DRL is expected as a promising alternative for automated trading in the metaverse ecosystem since DRL can enable the well-trained agent to make the decision automatically. 
Liu \textit{et al.} \cite{liu2021finrl,liu2021finrl2} believe that proper usage of AI will initiate a paradigm shift from the conventional trading routine to an automated machine learning approach. Therefore, Liu \textit{et al.} \cite{liu2021finrl} propose a DRL-based system to achieve efficiently automate trading in the ecosystem, named FinRL, which can solve dynamic decision-making problems and build a multi-factor model. In addition, Liu \textit{et al.} \cite{liu2021finrl2} attempt to reduce the simulation-to-reality gap and data processing burden through an open-source library that includes hundreds of market {\color{black}environments} for financial reinforcement learning.

\section{Blockchain in Metaverse}\label{sec:blockchain}

\begin{figure}
    \centering
    \includegraphics[width=0.5 \textwidth]{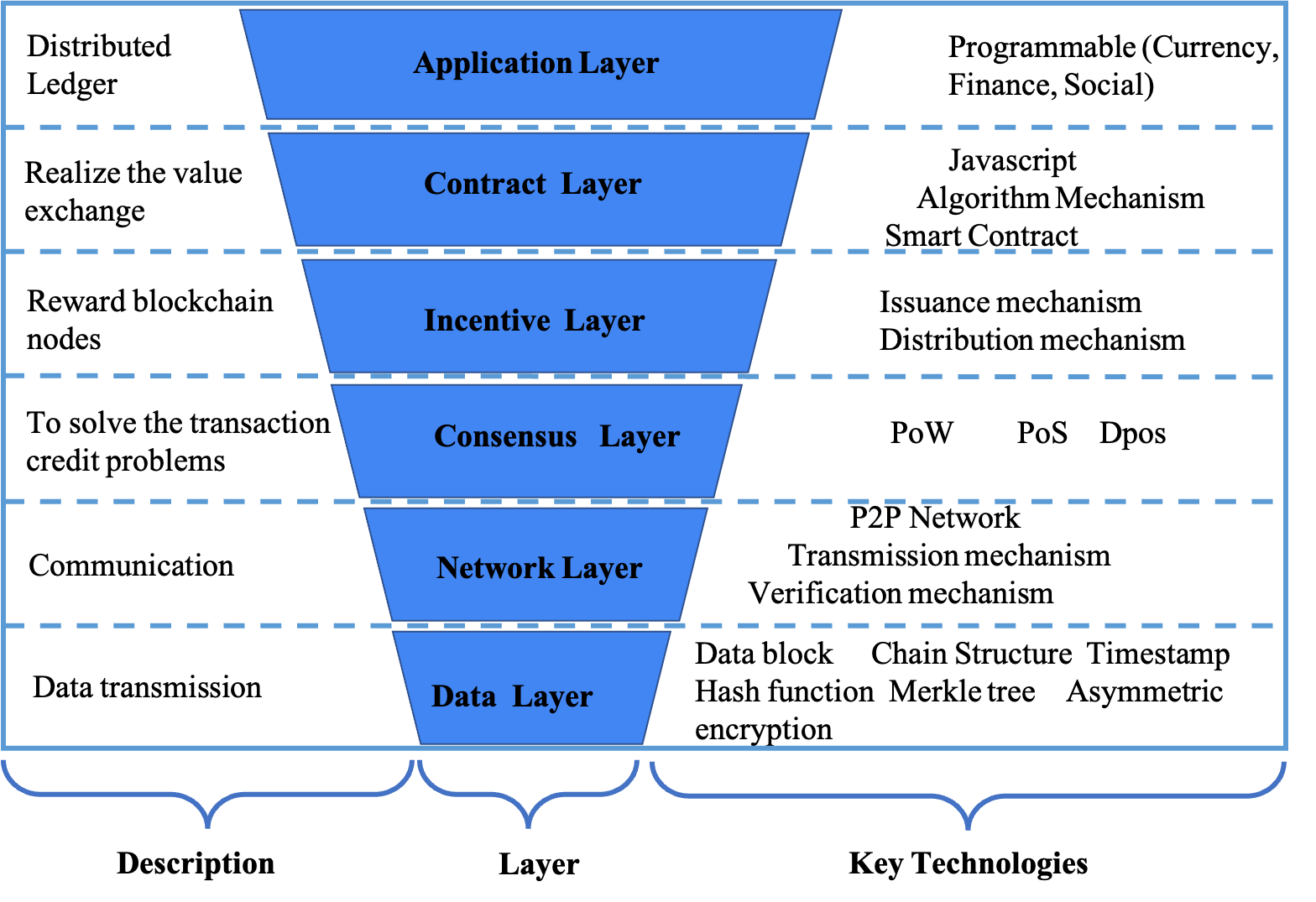}
    \caption{Layered Architecture of Blockchain.}
    \label{lbl:BCModelArchitecture}
\end{figure}

{\color{black}

Blockchain is widely believed as one of the fundamental infrastructures of metaverse, because it can bridge isolated small sectors together and provide a stable economic system, which helps offer transparent, open, efficient, and reliable rules for metaverse.
{\color{black}
 For example, hash algorithms and timestamp technologies as the major components in the data layer of blockchain could provide metaverse users the traceability and confidentiality of the data storing in the bottom layer of blockchains.
 }
As illustrated in Fig. \ref{lbl:BCModelArchitecture}, the conventional Blockchain architecture includes a data layer, network layer, consensus layer, incentive layer, contract layer, and application layer. The correlations between those layers and the metaverse are explained as follows.
\begin{itemize}

\item Data transmission and verification mechanism provide network support for various data transmission and verification of metaverse economic system.

\item Consensus mechanisms solve the credit problem of metaverse transactions. 

\item Distributed storage of blockchain ensures the security of virtual assets and the identities of metaverse users.

\item Smart contract technology offers a trustworthy environment for all participating entities in the metaverse. It realizes the value exchange in the metaverse and ensures the transparent execution of system rules described in contract codes. Once deployed, the code of smart contracts cannot be modified anymore. All clauses depicted in those smart contracts must be completely executed.

\end{itemize}

Without the support of blockchain technology, it will be difficult to identify the value of the resources and goods trading in the metaverse, especially when those virtual elements have economic interactions with the real-world economy.
Thus, it is undoubtedly worth exploring the blockchain technology in the metaverse. 

In this section, we introduce the blockchain-empowered applications in metaverse considering four perspectives, i.e., \textit{Cryptocurrency}, \textit{Transaction characteristics}, \textit{Authentication}, and \textit{Market \& Business}.

}

\subsection{Cryptocurrency for Metaverse}

{\color{black}

What is the functionality of cryptocurrency in the metaverse?  How to issue and apply cryptocurrency in the metaverse?
Faced with these questions, we discuss whether the conventional cryptocurrency-issuing rule applies to metaverse in this subsection.


Cryptocurrency is one of the main applications under the spotlight empowered by blockchain. It also makes blockchain more popular. The trust of a wide range of users supports the value system of cryptocurrencies and drives both the circulation and trading of cryptocurrencies. 
To date, more than 12,000 virtual currencies have been issued worldwide, and new virtual currencies are still being created every single day. The price of mainstream cryptocurrencies keeps hitting record highs. At the time of the writing, the exchange rate of the US dollar to Bitcoin hits 50000:1 (USD/BTC) recently.

 Like the real world builds upon fiat currencies, the future metaverse inevitably needs cryptocurrencies, which deliver value during their circulation, payment, and currency of settlement. In detail, blockchain systems have implemented a series of operations for cryptocurrencies, such as creation, recording, and trading. All these fundamental operations are necessary for the metaverse.

Traditionally, Bitcoin \cite{bit} adopts the UTXO (Unspent Transaction Outputs) transaction model to trace the usages of this cryptocurrency, while Ethereum \cite{ethereumw} records the balance of each account address, which can be queried directly through the Ethereum dataset tools (e.g. Etherscan). 
Both UTXO and Ethereum adopt the Proof of Work (PoW) consensus. Miners mint coins by generating new blocks. However, there is a cost for them to generate blocks. In such PoW consensus, miners mine blocks by calculating a hash puzzle, which consumes a giant amount of electricity. For Ethereum 2.0, miners under the PoS (Proof of Stake) consensus mechanism mine blocks by electing, which depends on the coin age of miners \cite{duong2018twinscoin}.

Using blockchain technologies, there are multiple ways to deal with cryptocurrency exchanges. The vast majority of cryptocurrency exchanges occur in centralized exchanges such as OKEx and AOFEX. The advantages of the centralized exchanges include low-latency transactions, simple interfaces and a certain level of security. However, the centralized exchanges also experience scandals such as price manipulation by insiders taking advantage of information asymmetries.
Other cryptocurrency exchanges occur in decentralized exchanges, where smart contracts or other peer-to-peer network execute transactions automatically \cite{daian2020flash}. In some cases, the smart contracts, e.g., IDex and Paradex, maintain continuous-limited order books offchain, a counterparty of the order or the exchange itself performs order matching and submits order pairs to the smart contract for processing. In other cases, such as Uniswap and Bancor, the smart contract performs as a counterparty and trade with its user directly.
%
%
It is easy to foresee that metaverse built by different corporations will coexist in the near future. Thus, various cryptocurrencies used in those smaller metaverses need to be exchanged like fiat currencies in the physical world. We envision that multiple cryptocurrencies will also coexist in future metaverse. The metaverse users will naturally need to exchange different cryptocurrencies in the same way as described above.
}

\subsection{Transaction Characteristics in Metaverse}

{\color{black}

 Metaverse is expected to have various types of financial issues such as estate purchases, item rentals and service acquisitions, which include almost everything people do in the physical world. Therefore, transactions in metaverse are not merely related to intra-metaverse scenario, and not only for token transferring.

 Once a user launches a transaction on the conventional blockchain, the transaction will be first broadcast to miners and be stored in their local transaction pools. The miner picks up a certain number of transactions and next performs the hash-based consensus.
 The block generated by the first miner to find an output to the puzzle that fits the specified difficulty will be uploaded to the chain and broadcast to all the other miners. 
 
 
 In metaverse, it is not hard to predict that those blockchain nodes will process a giant volume of transactions since metaverse has a significant number of users, who trade in the virtual world every second while using various intra- or inter-metaverse applications. 
 As conventional, the full blockchain nodes working in metaverse need to store all the historical transactions locally, which brings a great burden for the full nodes.

 
 Another issue associated to metaverse transactions is the requirement of low confirmation latency.
 Internet applications that satisfy human habits often have end-to-end latency between tens of milliseconds and hundreds of milliseconds. Even further, metaverse applications based on three-dimensional display and interaction require an latency within 10 milliseconds in order to avoid dizziness. All those low-latency applications enforce metaverse transactions to have low-confirmation latency.

 The existing blockchain consensus protocols have several constraints that prevent them from being directly applied to metaverse. 
 For instance, the PoW mechanism relies on miner's hash power to achieve consensus on a certain transaction data. 
 Considering the huge data volume of metaverse, PoW consensus will consume a large amount of mining resources. 
 The PoS mechanism, on the other hand, relies on the number and age of coins held by miner nodes for reaching consensus towards a group of proposed transactions. 
 Since the Matthew effect is more pronounced on metaverse, the PoS mechanism cannot ensure the fairness of the miners participating in the consensus in metaverse.
 Therefore, metaverse needs new consensus protocols and new blockchain mechanisms to meet the rigorous requirements of transactions.


 We then review some state-of-the-art studies to find some clues to address the transaction issues aforementioned.
  In \cite{ChainSplitter}, the authors merge blockchain with IIoT architecture and propose a hierarchical storage structure where historical information is stored on the cloud and the latest block is stored on IIoT devices. The new architecture could offer immutable and verifiable services. 
 To reduce the cost of block validation during forwarding, Frauenthaler \textit{et al.} \cite{Frauenthaler2020ETHRelay} propose to validate the block header only when needed, making interoperation between Ethereum-based blockchains feasible.
 Yang \textit{et al.} \cite{CoDAG} change the traditional linear structure of blockchains using Directed Acyclic Graph (DAG) structure where blocks are organized into a compacted DAG structure. This new structure can improve security and decrease the transaction verification time.
 Aumayr \textit{et al.} \cite{VirtualChannel} present a virtual channel protocol based on UTXO that is compatible with almost all cryptocurrencies, aiming to reduce the confirmation latency in the context of many transactions waiting to be processed. 
 To accelerate transaction relay in blockchain networks, Zhang \textit{et al.} \cite{zhang2021accelerating} propose a Repulay protocol where nodes select neighbors based on a reputation mechanism and verify transactions only with a certain probability.
Overall, the proposed RepuLay also helps nodes save their bandwidth while guaranteeing the quality of transaction relay.

}

\subsection{Blockchain-empowered Market in Metaverse}

 {\color{black}
 
 Blockchains before Ethereum, such as Bitcoin, only support token transferring. Until the emergence of the Ethereum platform, smart contracts begin to support Turing-complete programming. Complicated businesses could be executed in a virtual machine through smart contract codes. 
 Ethereum realizes the upgrade of blockchain applications from cryptocurrency to crypto-business. Various blockchain reconfiguration of market and business could be implemented. 
 
 Empowering by the advanced blockchain technologies, decentralized finance (DeFi) can boost the decentralized market and business in the metaverse. 
 We review several representative studies related to the DeFi market and business here.
 In \cite{daian2020flash}, the authors analyze the behavior of arbitrage bots in the context of the cryptocurrency market. They find that arbitrage robots could observe the transactions in the transaction pool and perform arbitrage without risks. They also present a cooperative strategy to maximize the profit of arbitrage robots and point out that miners could act as arbitrage robots under certain circumstances. However, the MEV (miner extractable value) could incentivize the emergence of forking attacks.
 The authors propose a cooperative bidding strategy for them to strive for more profit. 
 They also find that the current amount of MEV in a month is more than 25$\times$ the cost of a 51\% attack on Ethereum.
 DeFi, based on smart contracts and Fungible Tokens (FT), offers a new approach to innovate economic models in the metaverse. Existing successful solutions, such as Uniswap \cite{angeris2019analysis}, a decentralized exchange (DEX) implemented on Ethereum, automatically provide users with liquidity for their tokens. 
 DEXs is a new kind of marketplace that can offer a secure peer-to-peer exchange of crypto-asset tokens for trading \cite{dai2020dex}. The core of a DEX, named \textit{atomic swap}, enables two parties to exchange tokens or crypto-assets without involving an intermediary party.
 Cybex \cite{cybex} as a DEXs-based DApp, provides a peer-to-peer marketplace for tokens to be exchanged. Cybex also issues its token called CYB. 
 It is worth noting that CYB is only allowed to use in the Cybex market when paying for the exchange of new tokens, staking to borrow crypto-asset tokens, paying transaction fees, etc.
 The Diem Blockchain \cite{diem} is the technological backbone of the payment system, operated by a network of validator nodes. The software that implements the blockchain is open-sourced so that anyone can build upon it and scale their financial needs.

\subsection{Blockchain-empowered Authentication in Metaverse}

 Currently, the economic activities in metaverse mainly include the auction of virtual assets, including land, scarce items, precious real estate, the development and leasing of land, the rewards of game tasks, and the profits from investing in cryptocurrency. Thus, metaverse invokes a new form of funding that draws inspiration from both the real and virtual world. 

 NFT has been mainly used to commemorate special moments or to collect digital assets, and recently it is creating a new digital content business by combining it with metaverse \cite{Ethereum,MetaerseP}.
 NFT can guarantee the uniqueness of the digital assets by keeping encrypted transaction history permanently on the blockchain. Each token has a unique recognizable value, which enables to authenticate the ownership of digital assets. 
 For example, the blockchain-empowered NFT has been applied to prove the uniqueness of the avatar and the created things in metaverse \cite{jeon2021blockchain}.
 }
 
\begin{table*}
\caption{A Summary of the Blockchain Technologies in Metaverse}
\center
\footnotesize
\setlength\tabcolsep{1.5pt} 
\begin{tabular}{|p{2.5cm}|p{7.0cm}|p{3cm}|p{5cm}|}
\hline
\bf{ Types} & \bf{Description} & \bf{Representative Works}  & \bf{Use cases}  \\ \hline \hline
\multirow{2}{*}{\shortstack[l]{Blockchain \\ Architecture}} & Smart contracts & \cite{daian2020flash} &  Executing transactions  \\ \cline{2-4}
&Consensus mechanisms  & \cite{duong2018twinscoin} &  PoW, PoS \\
\hline

\multirow{7}{*}{\shortstack[l]{Applications of \\ Blockchain}} & Cryptocurrency & \cite{bit,ethereumw} &Bitcoins, Ethereum \\ \cline{2-4} 
&  Repulay protocol based on reputation mechanism & \cite{zhang2021accelerating}  & Accelerate transaction relay \\ \cline{2-4}
& Incorporate probabilistic blockchain with reputation & \cite{salman2019reputation}  & Blockchain and AI\\ \cline{2-4}
& Alliance chain with reputation & \cite{ malik2019trustchain}  & Supply chain \\ \cline{2-4}
& Blockchain Meets IoT & \cite{novo2018blockchain}  & Use Blockchain to manage IoTs \\  \cline{2-4}
 & Fast certificate verification& \cite{wang2020collaborative}  & Certificate verification \\ 
\hline 

\end{tabular}
\label{lbl:block}
\end{table*}

\section{Integration of Blockchain and AI on Metaverse}

As aforementioned, we have discussed the details of AI and blockchain, and their impacts on the metaverse. The integration of AI and blockchain, namely \textit{blockchain intelligence} and \textit{intelligent blockchain}, shall be explored due to their close interactions.
For example, the decentralized AI as an integration of AI and blockchain \cite{ai2018decentralized} enables to process and perform analytics or decision making on trusted data without any support from trusted third parties.
Hence, in this section, we shall discuss their integration and seek to answer the question: what contributions can intelligent systems make in metaverse \cite{8481263}.

\subsection{Blockchain for AI}

Blockchains can offer various components for AI, including dataset, algorithms, and computing power through trading from the decentralized marketplace. Blockchain encourages the innovation and adoption of AI to an unprecedented level in the context of metaverse.
Several representative examples are reviewed as follows.
Mamoshina \textit{et al.} \cite{mamoshina2018converging} propose a blockchain-based decentralized model that enables users to access their personal data in an AI-moderated healthcare data exchange. 
The proposed model allows users to upload their data directly to the system and grants access to their data using transparent pricing. Such transparent pricing is determined by a data value model, guaranteeing fair tracking of all data usage activities.
Woods \cite{woods2018blockchain} analyzes the importance of fusing AI techniques and blockchain infrastructure, aiming to address the security risks faced by the Internet. The author finds that bots-bots and human-bots interactions have increased since 52\% of the web traffic is generated by bots. Thus, bot-bot communications will exceed human-bot interactions according to the increasing bot traffic. In metaverse, each person will have multiple avatars. This fact inevitably leads to huge traffic due to massive interactions. 


\subsection{AI for Blockchain}

To design a blockchain, developers have to tune massive parameters and consider trade-offs between incentive, consensus, security, and many other aspects. 
AI technologies can be applied to deal with those problems and help blockchain systems achieve higher performance.
In addition, with the breakthroughs of machine learning, a blockchain governed by a machine learning-based algorithm might enable automatically detection of attacks and invoke appropriate defense mechanisms. 
For example, Salimitari \textit{et al.} \cite{9013824} propose a machine learning-based framework, which aims to offer a secure and robust consensus in blockchain-based IoT networks. The authors then present a two-stage consensus protocol for the AI-enabled blockchain that exploits an outlier detection algorithm in an IoT network.


\section{Challenges and Open Issues} \label{sec:challenges}

Through the previous review, we found that AI and blockchain are fundamental technologies for the metaverse. Although those technologies are promising to build a scalable, reliable, and efficient metaverse, we are aware that the metaverse is still in its infant stage. Thus, this section discusses the challenges, open issues, and suggestions for the fusion of AI and blockchain in the metaverse.

\subsection{Open Issues on Digital Economy in Metaverse}

{\color{black}

Different from the physical world, the digital creation in the virtual world might be unlimited. The identity of digital objects determines value instead of the undifferentiated labor in the conventional economy.
In the field of digital creation, it is necessary to develop authoring tools to enable the users to produce original content easily and gain rewards efficiently at {\color{black}a} low cost. Those tools could improve the enthusiasm of content producers of the metaverse. The marginal benefits will increase in the metaverse instead of diminishing marginal benefits of production in the physical world. The difference in marginal benefits between the physical world and the virtual world demands a value conversion mechanism to bridge their gap.

In future metaverse, people prefer to turn to their virtual cabinet to select a digital outfit, while companies begin to hype the virtual skins, virtual clothing, and even virtual estates with a high price which will block a large portion of players to join in the metaverse. 
Hence, it is necessary to propose particular governance mechanisms under the cooperation of worldwide companies.
Furthermore, how to establish a digital currency system that enables the currency exchange between the metaverse and the physical world remains an open issue.

}

{\color{black}

%
In addition, the transaction volume and frequency that occurred in the metaverse will become extremely much higher than happened in the physical world.
%
%
Thus, how to support such high-volume and high-frequency transactions remains a challenging problem in the future metaverse.
%
%
%
Another issue related to future metaverse might be the inflation caused by massive cryptocurrency supplements in a decentralized economy system built upon blockchain and AI technologies.

}
 
\subsection{Artificial Intelligence Issues}

 The breakthroughs of artificial intelligence technologies, especially deep learning, enable  academia and industries to make great progress in the automatic operation and design in the metaverse and perform better than conventional approaches. 
 For example, the study \cite{Epic} applies AI to generate vivid digital characters quickly that might be deployed by virtual service providers as conversational virtual assistants to populate the metaverse.
 However, existing deep learning models are usually very deep and have a massive amount of parameters, which incurs a high burden for resource-constrained mobile devices to deploy learning-based applications.
While current AI technologies are just at the stage where people tell the machine to do specific tasks instead of enabling the machine to learn to learn automatically. 
Most learning tasks are only suitable for the closed static environment and have poor robustness, and poor interpretability that can not satisfy the requirement of availability, robustness, interpretability, and adaptability in an open and dynamic environment.
 

 %
 Meta-learning \cite{vanschoren2018meta} is a promising learning paradigm {\color{black}that} can observe how different machine learning approaches perform on a wide range of learning tasks. 
 Learning from this experience, meta-data can learn new tasks much faster than others possible.
 Not only does meta-learning dramatically speed up and improve the design of machine learning pipelines or neural architectures, it also allows us to replace hand-engineered algorithms with novel approaches learned in a data-driven way. 
 Therefore, meta-learning remains challenging to achieve auto-machine learning in future years.

\subsection{Blockchain-related Issues}

Although blockchain technologies have achieved a lot of improvements, there are still challenges and open issues while fusing blockchain in metaverse. We post several questions in the following to inspire readers to deeply dive into the related technical studies.

\begin{itemize}
 \item Can the existing real-world NFT {\color{black}platforms} adapt to the high transaction volumes in the metaverse? 
 {\color{black} NFTs are unique cryptographic tokens that deployed on blockchains. Assigning non-reproducible features, NFTs digitalize real-world items like artworks and real estate. 
 However, the current NFT platforms are in their initial stage. To meet the high-volume requirement of future metaverse's applications, to improve the service level of NFT platforms is an essential research and engineering topic.
 }

 \item What rules does metaverse require towards a healthy digital blockchain-empowered market and business?
 {\color{black}
 From the perspective of policy, the combination of decentralization and regulation might be a promising way for the digital blockchain-empowered market.
 DAO \cite{liu2021dao} is regarded as an efficient, decentralized, and promising paradigm that works with like-minded folks around the globe. The members in DAO have built-in treasuries that no one has the authority to access without the approval of the group. 
 DAOs are executed through smart contracts, which are transparent codes verifiable by anyone.
 Thus, exploiting DAOs, decisions can be governed by proposals and voting, to ensure everyone in the metaverse has a countable voice. 
 
 }
  
 \item Is the real-world blockchain-empowered application model able to be directly transplanted to metaverse? 
 {\color{black} 
 Nowadays, the overwhelming speculations over thousands of cryptocurrencies and the scams of \textit{initial coin offering} (ICO) have brought notorious doubts on metaverse. 
 Moreover, the existing blockchain-empowered application models cannot meet the stringent requirements of metaverse with it low latency and high throughput performance. 
 Hence, a high-performance and secure blockchain-empowered application model is needed for the various applications used in future metaverse. 
 }
 
 \item Does metaverse need new blockchain platforms and new consensus mechanisms? {\color{black} 
 The foundation of the blockchain is consensus mechanism, such as PoW and PoS.
 However, the existing consensus mechanisms have shortcomings with regard to large amount of hash computing and various security issues.
 %
 %
 The future decentralized applications demand a blockchain platform to fulfill the following desired characteristics, i.e., low latency, high throughput, fast transaction-sequential performance, offline payment of transactions, low transaction fee, modern free internet business model, etc. Hence, new blockchain platforms and new consensus mechanisms are expected to appear for future metaverse.
 }
 \end{itemize}

\subsection{Governance in Metaverse}

 Currently, the concept of the metaverse is mainly used and propagated by companies such as Roblox and Meta (Facebook previously). Predictably, the most popular ecosystems soon are built and operated by these giant corporations. 
 Tiny metaverse has only a few application scenarios. In contrast, the macro metaverse would include all scenarios required by users. 
 To realize such a vision, large companies need to cooperate to create a huge unified metaverse.
 The problem is how to incentivize those giant companies to cooperate? Once the unified metaverse is set up, how do make uniform rules that govern the whole unified metaverse? 
 
 On the other hand, the threats of market manipulation and money laundering will exist in the future metaverse. Thus, the market governance will be viewed as more significant from the perspective of regional jurisdiction of the metaverse.




\subsection{Blockchain-empowered Applications for Metaverse}

Various applications will boost the virtual economy in the metaverse, such as blockchain-empowered Apps for office work, social networks, NFT markets, game finance, etc. 
For instance, the blockchain-based game Axie Infinity \cite{Axie} establishes a digital pet universe in which players can battle, raise, and trade fantasy creatures called Axies. The game allows players to deposit from an Ethereum wallet to a Ronin wallet via the Ronin Bridge. In short, the game allows the currency exchange between fiat and cryptocurrency since the players can buy ETH on a cryptocurrency exchange like Binance or Coinbase or with fiat on Ronin, and send it to their address due to the legality of cryptocurrency in some countries. %
Although this blockchain-empowered game finance does not work in some countries, we believe that the future metaverse will embrace a much more open, fair, and rational physical world.

\subsection{Security and Privacy for Metaverse}

From the perspective of metaverse companies, developers, and metaverse users alike, a natural question is how to guarantee their security \& privacy in metaverse which could mean violation of their privacy, potential identity theft, and other types of fraud \cite{dataPrivacy}. 

For example, a lot of private properties in the metaverse, including the digital assets, the identity of virtual items, cryptocurrency spending records, and other private user data, are required to be protected. Thus, metaverse-oriented cryptography mechanisms are open proposals for privacy preservation in the metaverse.

\section{Conclusion}\label{sec:conlusion}


Artificial intelligence and blockchain technologies are expected to play essential roles in the ever-expanding metaverse.
For example, metaverse uses artificial intelligence and blockchain to create a digital virtual world where anyone can safely and freely engage in social and economic activities that transcend the limits of the real world. 
Exploiting metaverse, the application of these latest AI and blockchain technologies will be accelerated as well.

By surveying the most related works across metaverse components, digital currencies, AI technologies and applications in the virtual world, and blockchain-empowered technologies, we wish to offer a thoughtful review to the experts from both academia and industries.
We also envisioned critical challenges and open issues in constructing the fundamental elements of metaverse with the fusion of AI and blockchain.
Further exploitation and interdisciplinary research on the metaverse entail {\color{black}collaboration}  from both academia and industries to strive for an open, fair and rational future metaverse.
\normalem
 \bibliographystyle{unsrt}
\bibliography{references}  


\end{document}